\documentclass[a4paper]{article}
\usepackage{graphicx}
\usepackage{twocolceurws}
\usepackage{hyperref}
\usepackage{array}
\title{Learning Emoji Embeddings using Emoji Co-occurrence Network Graph}

\author{
Anurag Illendula \\ Department Of Mathematics\\
                IIT Kharagpur \\ aianurag09@iitkgp.ac.in
\and
Manish Reddy Yedulla \\Department of Engineering Science\\
IIT Hyderabad \\ es15btech11012@iith.ac.in
}

\institution{}

\begin{document}
\maketitle

\begin{abstract}

Usage of emoji in social media platforms has seen a rapid increase over the last few years. Majority of the social media posts are laden with emoji and users often use more than one emoji in a single social media post to express their emotions and to emphasize certain words in a message. Utilizing the emoji co-occurrence can be helpful to understand how emoji are used in social media posts and their meanings in the context of social media posts. In this paper, we investigate whether emoji co-occurrences can be used as a feature to learn emoji embeddings which can be used in many downstream applications such sentiment analysis and emotion identification in social media text. We utilize 147 million tweets which have emojis in them and build an emoji co-occurrence network. Then, we train a network embedding model to embed emojis into a low dimensional vector space. We evaluate our embeddings using sentiment analysis and emoji similarity experiments, and experimental results show that our embeddings outperform the current state-of-the-art results for sentiment analysis tasks.

\end{abstract}

\section{Introduction}

Emojis are the 21st century\textsc{\char13}s successor to the emoticon. They arose from the need to communicate body language and facial expressions during text conversations. They are two-dimensional visual embodiments of everyday aspects of life which were standardized by the Unicode Consortium in 2010 as part of Unicode 6.0. Emoji proliferated throughout the globe and has particularly become a part of the popular culture in the west. It has been adopted by almost all social media platforms and messaging services. Emojis serve many purposes during online communication, among which conveying emotion is one of the primary uses. According to the latest statistics released by Emojipedia in June 2017, the number of emojis has increased to 2,666, posing challenges to applications that list them in small hand-held devices such as mobile phones. To overcome this challenge, emoji keyboards in most of the smartphones contains categorizes emoji into several categories listed in \hyperref[sec:tableofemojis]{Table 1}.  

Many recent Natural Language Processing (NLP) systems rely on word representations in finite-dimensional vector space. These NLP systems mainly use pre-trained word embeddings obtained from word2vec~\cite{mikolov2013distributed} or GloVe \cite{pennington2014GloVe} or fastText \cite{bojanowski2016enriching}. Earlier GloVe embeddings were used for training most NLP systems, but fastText trained word embeddings could achieve much higher accuracies of NLP systems involving social media data because the fastText model could learn sub-word information. Emoji embeddings have been of fundamental importance to improve the accuracies of many emoji understanding tasks. Recent research proved that emoji embeddings could enhance the performance of emoji prediction~\cite{felbo2017using,barbieri2017emojis}, emoji similarity~\cite{wijeratne2017semantics}, and emoji sense disambiguation tasks~\cite{emojineticwsm, sheth2017knowledge}. These emoji representations have also been efficient in understanding the behavior of emojis in different contexts. The need to learn emoji representations for improving the performance of social NLP systems has been recognized by Eisner et al. \cite{eisner2016emoji2vec} and Francesco et al. \cite{barbieri2016does} among others, where they used traditional approaches which include skip-gram and CBOW model to learn emoji embeddings. 

Information networks such as publication networks, World Wide Web are characterized by the interplay between various content and a sophisticated underlying knowledge structure. Graph embedding models are helpful to scale information from large-scale information networks and embed them into a finite-dimensional vector space, and these embeddings have shown great success in various NLP tasks such as node classification \cite{bhagat2011node}, link prediction \cite{liben2007link} and classification \cite{yu2014personalized} tasks. These graph embedding models have been of crucial importance and have enhanced the performance of word similarity and word analogical reasoning tasks using language networks \cite{tang2015line}. The analysis of emoji co-occurrence network graphs can help us understand emojis from different perspectives. We hypothesize that emojis which co-occur in a tweet contains the same sentiment as the overall sentiment of the tweet. Consider a tweet, ``I got betrayed by \includegraphics[height=1em]{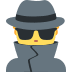}, I want to kill you \includegraphics[height=1em]{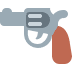}'', here both the emojis \includegraphics[height=1em]{Emojis/2024.png},\includegraphics[height=1em]{Emojis/1281.png} contain negative sentiment, the overall sentiment of the tweet is also negative. Hence we investigate whether emoji co-occurrence could be a better feature to learn emoji representations to improve the accuracy of classification tasks. In this paper, we introduce an approach to learn emoji representations using emoji co-occurrence network graph and large-scale information network embedding model and evaluate our embeddings using the gold-standard dataset for sentiment analysis task.

\begin{table}[ht]
\begin{center}
\caption{Emoji Categories}
\label{sec:tableofemojis}
\bigskip

\begin{tabular}{| >{\centering\arraybackslash}m{1in} | >{\centering\arraybackslash}m{1in} |}
\hline
Category & Emoji Examples \\
\hline
Smiley and People & \includegraphics[height=1em]{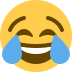},\includegraphics[height=1em]{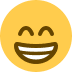},\includegraphics[height=1em]{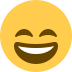} \\ 
Animals and Nature & \includegraphics[height=1em]{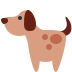},\includegraphics[height=1em]{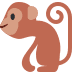},\includegraphics[height=1em]{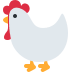} \\
Food and Drink & \includegraphics[height=1em]{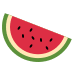},\includegraphics[height=1em]{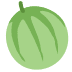},\includegraphics[height=1em]{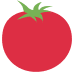}\\ 
Activity & \includegraphics[height=1em]{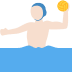},\includegraphics[height=1em]{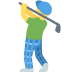},\includegraphics[height=1em]{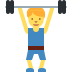}\\
Travel and Places & \includegraphics[height=1em]{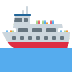},\includegraphics[height=1em]{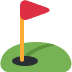},\includegraphics[height=1em]{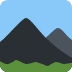}\\
Objects & \includegraphics[height=1em]{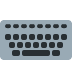},\includegraphics[height=1em]{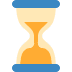},\includegraphics[height=1em]{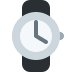}\\
Symbols & \includegraphics[height=1em]{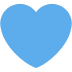},\includegraphics[height=1em]{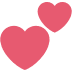},\includegraphics[height=1em]{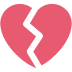}\\
Flags & \includegraphics[height=1em]{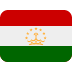},\includegraphics[height=1em]{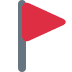},\includegraphics[height=1em]{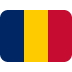}\\ \hline
\end{tabular}
\end{center}
\end{table}

\begin{figure}
\label{sec:emojidistribution}
\centering
\includegraphics[width=1.0\linewidth]{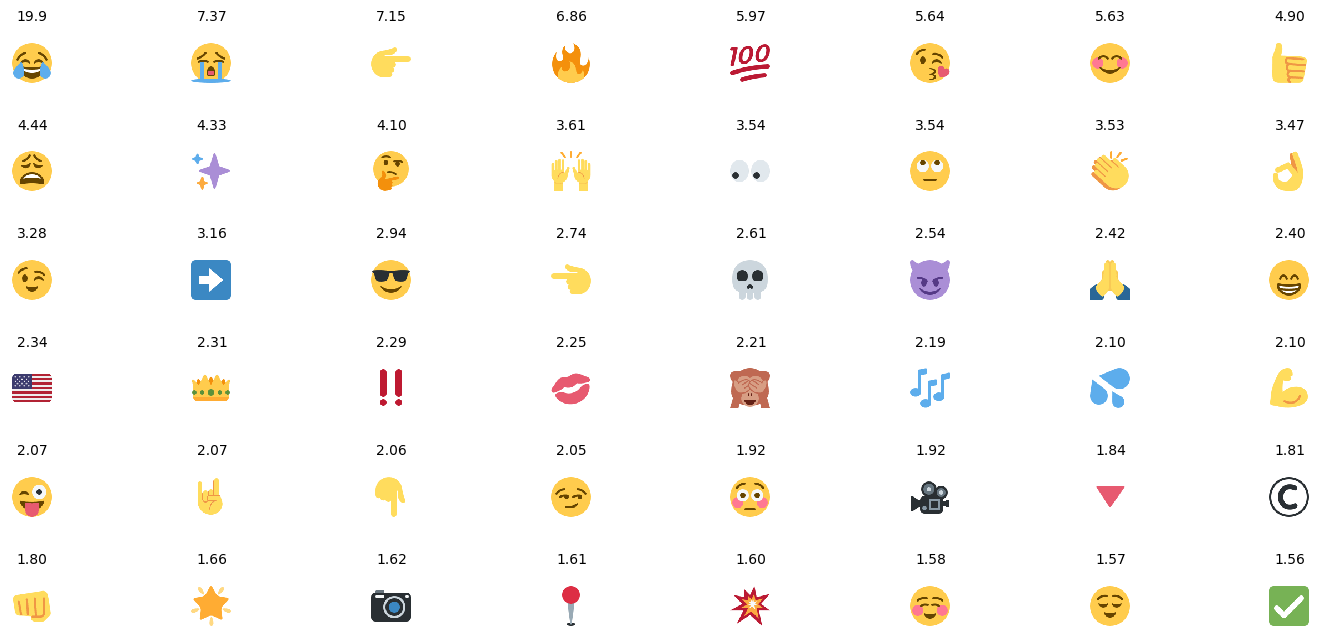} 
\caption{Distribution of Tweets across various emojis in lakhs}.
\label{fig1}
\end{figure}

This paper is organized as follows. Section 2 discusses the related work done by other researchers in the field of emoji understanding and learning network representations. Section 3 discusses the process of creating an emoji co-occurrence network using our twitter corpus. Section 4 explains our model architecture to learn emoji representations from emoji co-occurrence network graph. Section 5 reports the accuracies obtained by our emoji embeddings on the gold-standard dataset for sentiment analysis task, emoji similarity tasks. We discuss the reason behind high accuracies obtained for sentiment analysis task in Section 6 followed by plans for future work in Section 7.

\section{Related Work}

One of the exciting work by Wijeratne et al. \cite{wijeratne2016emojinet,emojineticwsm} in the field of emoji understanding is EmojiNet (\href{http://emojinet.knoesis.org/home.php}{http://emojinet.knoesis.org/home.php}), the largest machine readable emoji sense inventory, this inventory helps computers understand emojis. In this work Wijeratne et al. tried to connect emojis and their senses to corresponding words in babelnet (\cite{NavigliPonzetto:12aij}) using their respective babelnetId. EmojiNet opened doors to many of the emoji understanding tasks like emoji similarity, emoji prediction, emoji sense disambiguation.

The other interesting work done by Wijeratne et al. \cite{wijeratne2017semantics} addressed the challenge of measuring emoji similarity using the semantics of emoji. They defined two types of semantics embeddings using the textual senses and the textual descriptions of emojis. Prior work by Francesco et al. (\cite{barbieri2016does}) and Eisner et al. (\cite{eisner2016emoji2vec}) used traditional approaches to learn emoji embeddings. The semantic embeddings have achieved accuracies which outperformed the previous state-of-the-art results in sentiment analysis task; this high accuracy is due to the fact that semantic embeddings can learn syntactic, semantic, sentiment features of emojis.

Seyednezhad et al. (\cite{seyednezhad2017understanding}) created a network using the emoji co-occurrences in the same tweet; they claim that each edge weight can help us understand the user's context to use multiple emojis. This emoji network also enabled them to justify the use of co-occurred emojis in different perceptions. This also enabled them to understand emoji usage by understanding possible relations between these special characters in common text. Fede et al. (\cite{fede2017representing}) studied different characteristics of this emoji co-occurrence network graph which include studying user's behavior to use a sequence of emojis in different contexts.

Information networks have been of primary use to store large amounts of information. Many researchers have proposed different graph embedding models in machine learning literature which allow us to embed nodes of large information networks into a low dimensional vector space (\cite{perozzi2014deepwalk} \cite{grover2016node2vec} \cite{cao2015grarep}). These embeddings helped address many tasks such as node classification, visualization, and link prediction tasks.

\begin{figure}
\label{sec:EmojiNetworkexamples}
\centering
\includegraphics[width=1.0\linewidth]{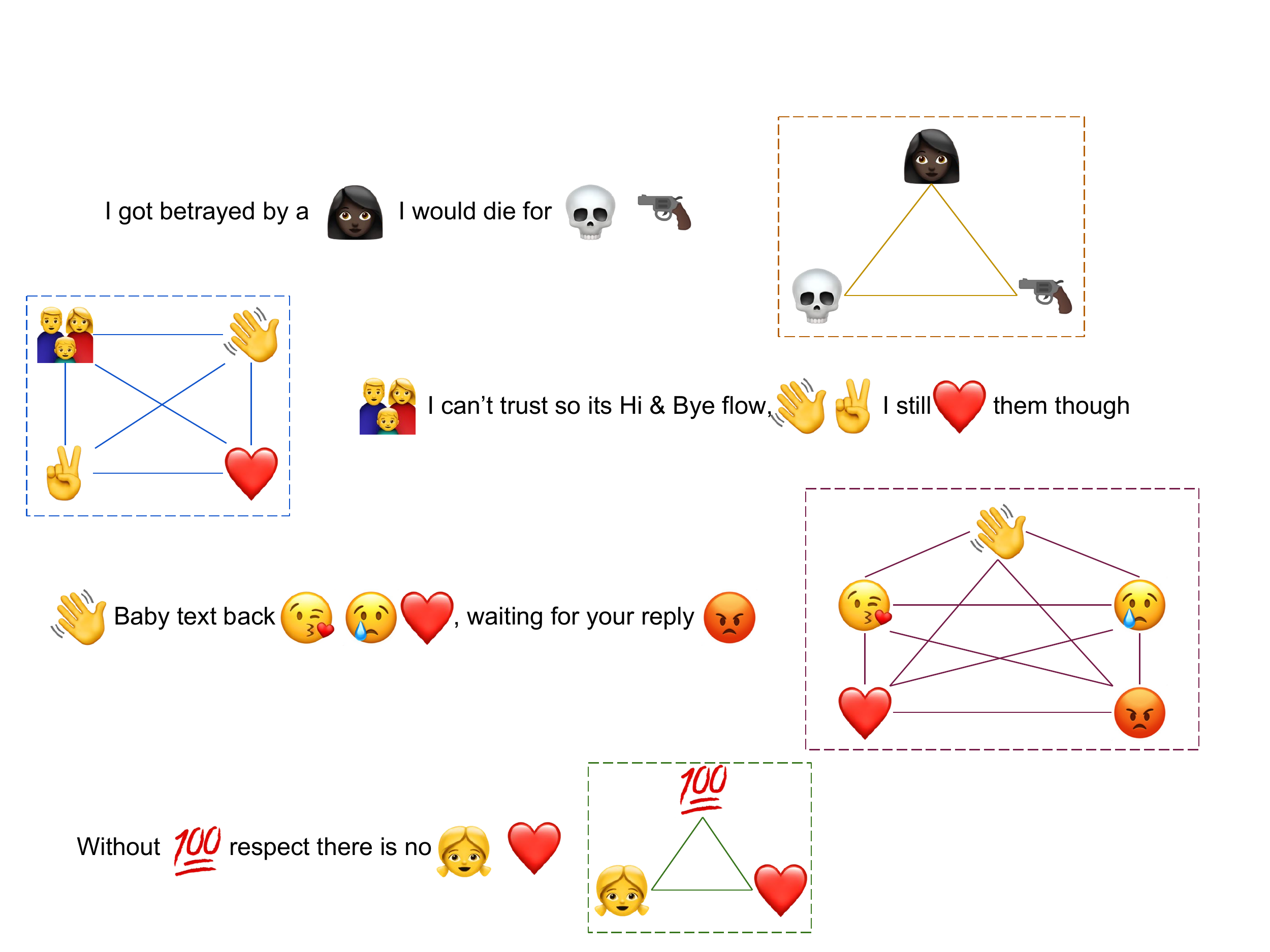} 
\caption{Construction of Emoji polygons}.
\label{fig2}
\end{figure}

\section{Data and Network}

The emoji network is constructed using a twitter corpus of 147 million tweets crawled through a period of 2 months (from $6^{th}$ August 2016 to $8^{th}$ September 2016) by Wijeratne {\em{et al.}}~\cite{emojineticwsm}. We filter the tweets and only consider the tweets which have multiple emojis embedded in a tweet. This reduces the number of distinct tweets in the dataset to 14.3 million. \hyperref[sec:emojidistribution]{Figure 1} shows the distribution of the number of tweets of the most frequently occurring emojis. Each tweet generates a polygon of $n$ sides where $n$ is the number of emojis embedded in the tweet. The construction of emoji network is straightforward and \hyperref[sec:EmojiNetworkexamples]{Figure 2} explains the construction of emoji polygons with the help of different examples. 

\begin{table}[ht]
\begin{center}
\caption{Most frequently co-occurring emoji pairs}

\bigskip
\label{sec:maxemojioccurences}
\begin{tabular}{| >{\centering\arraybackslash}m{1in} | >{\centering\arraybackslash}m{1in} |}
\hline
Emoji Pair & No of Co-occurrences \\
\hline
(\includegraphics[height=1em]{Emojis/991.png},\includegraphics[height=1em]{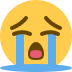}) & 230957 \\  (\includegraphics[height=1em]{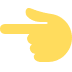},\includegraphics[height=1em]{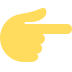}) & 196970 \\
(\includegraphics[height=1em]{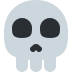},\includegraphics[height=1em]{Emojis/991.png}) & 135595 \\ 
(\includegraphics[height=1em]{Emojis/991.png},\includegraphics[height=1em]{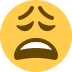}) & 102612\\
(\includegraphics[height=1em]{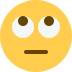},\includegraphics[height=1em]{Emojis/991.png}) & 102408\\
\hline
\end{tabular}
\end{center}
\end{table}

\begin{table}[ht]
\begin{center}
\caption{Least frequently co-occurring emoji pairs}

\bigskip
\label{sec:minemojioccurences}
\begin{tabular}{| >{\centering\arraybackslash}m{1in} | >{\centering\arraybackslash}m{1in} |}
\hline
Emoji Pair & No of Co-occurrences \\
\hline
(\includegraphics[height=1em]{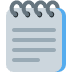},\includegraphics[height=1em]{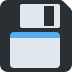}) & 1 \\
(\includegraphics[height=1em]{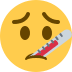},\includegraphics[height=1em]{Emojis/1053.png}) & 1 \\
(\includegraphics[height=1em]{Emojis/1053.png},\includegraphics[height=1em]{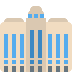}) & 1 \\ 
(\includegraphics[height=1em]{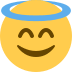},\includegraphics[height=1em]{Emojis/1053.png}) & 1\\
(\includegraphics[height=1em]{Emojis/1053.png},\includegraphics[height=1em]{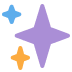}) & 1\\
\hline
\end{tabular}
\end{center}
\end{table}

\phantom{x}

The weight of an edge signifies the number of co-occurrences of the emojis sharing the edge considering the complete twitter corpus. For example in the case of tweets shown in \hyperref[sec:EmojiNetworkexamples]{Figure 2} the emoji pair (\includegraphics[height=1em]{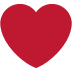},\includegraphics[height=1em]{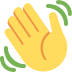}) appeared twice hence the weight of the edge corresponding to these two emojis is considered as $2$. Similarly, the weight of all the edges in the emoji network is calculated.

\begin{figure}
\label{sec:EmojiNetwork}
\centering
\includegraphics[width=1.0\linewidth]{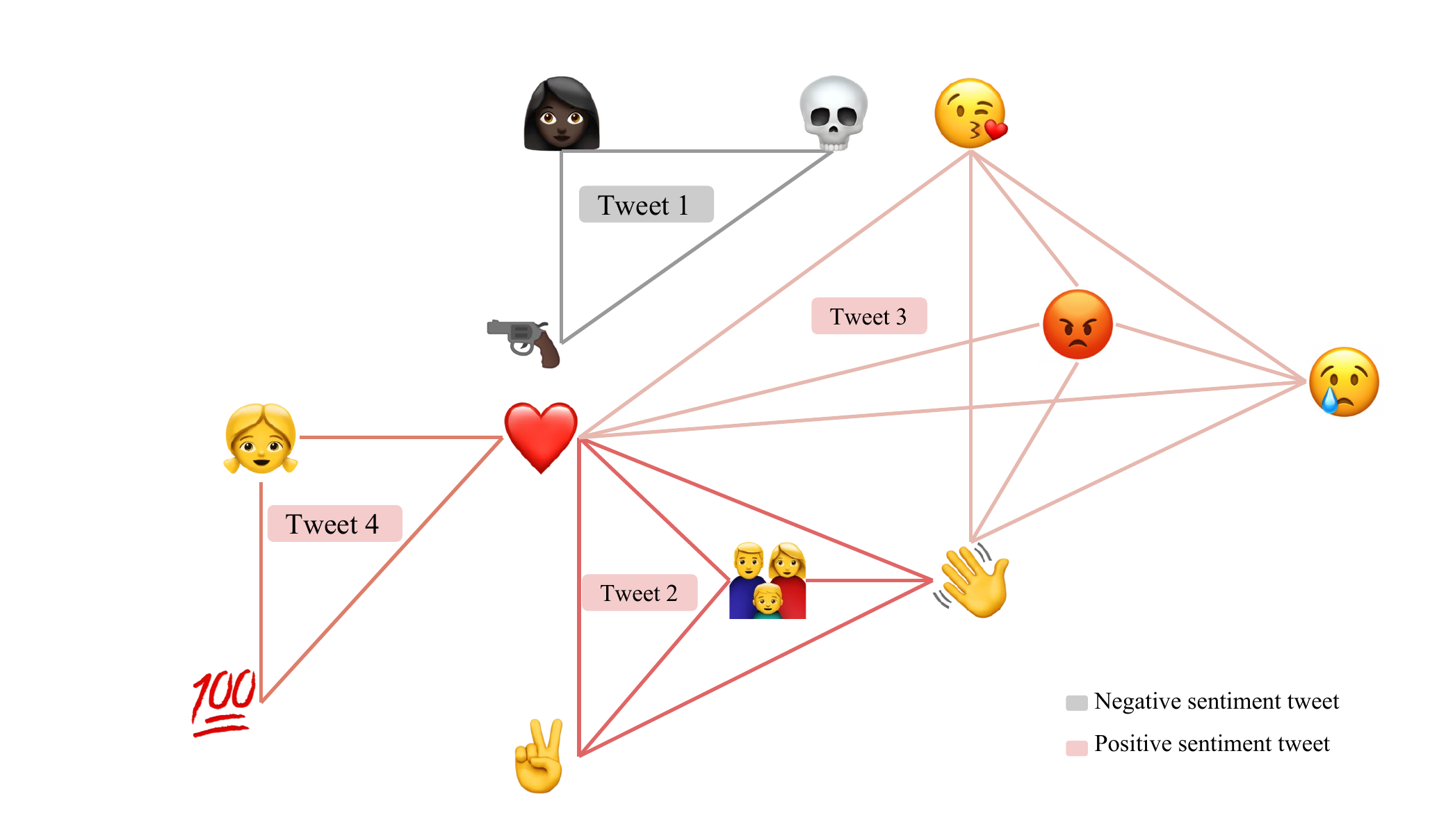} 
\caption{Construction of Emoji Network}.
\label{fig3}
\end{figure}

The emoji co-occurrence network created using the tweets in \hyperref[sec:EmojiNetworkexamples]{Figure 2} is represented in \hyperref[sec:EmojiNetwork]{Figure 3}. We input the emoji co-occurrence network graph to our graph embedding model to learn 300-dimensional emoji embeddings, and we evaluate our embeddings using the gold-standard dataset for sentiment analysis. We use the gold-standard dataset (\cite{novak2015sentiment}) to evaluate our embeddings because the current state-of-the-art results \cite{eisner2016emoji2vec} for sentiment analysis were obtained on this dataset. 

\section{Model}

\subsection{Description}

Here we discuss two different types of measures which signify the proximity between two nodes of the co-occurrence network graph, and the model developed by Jian et al. \cite{tang2015line} to learn the node representations of a network graph.

\phantom{x}

\textbf{First Order Proximity} : The first order proximity is defined as the local pairwise proximity which can be related to the weight of the edge formed by joining the two vertices. The first order proximity between an edge (u,v) is the weight $W_{uv}$ of the edge formed by vertices u, v. It can also be inferred from the definition that the first-order proximity between any two non-connected vertices is zero. 

\phantom{x}

\textbf{Second Order Proximity} : The second order proximity is defined as the similarity between neighbourhood network structures. For example, consider u to be an emoji node, let $p_u = (w_{(u,1)}, w_{(u,2)}, ....... ,w_{(u,|V|)})$  denote the first order proximity of the emoji node ``u'' with all the vertices then the second order proximity is defined as the similarity between $p_u$ and $p_v$. If there exists no common vertex between u and v, then second-order proximity is zero.

\subsubsection{Network embedding using first order proximity:}

Let $u_i$ and $u_j$ represent the network embedding in $d$ dimensional vector space, where (i,j) is an undirected edge in the network graph. The joint probability which signifies the proximity between vertices $v_i$, $v_j$ is defined as 

\begin{equation}
  p_1(v_i,v_j) = \frac {1} {1+exp(\vec{u}^T_i \cdot  \vec{u}_j)}  
\end{equation}

where $\vec{u_i} \in R^d$ is a low dimensional representation also called as embedding for emoji node $v_i$, $w_{ij}$ represents the weight of the edge between the nodes $v_i$ and $v_j$ The probability distribution between different pair of vertices is defined as p(.,.) over the vector space V x V and the empirical probability is defined as $\widetilde{p_1}(i,j)$

\begin{equation}
    \widetilde{p_1}(i,j) = \frac{w_{ij}}{W} \ \ \ and \ \ \ W = \sum_{(i,j) \in E}^{}w_{ij}
\end{equation}

To maintain the first Order proximity between the vertices of the network graph, the objective function ($O_1$) which is the distance between the empirical probability function and the proximity function is to be optimized.

\begin{equation}
  O_1(i,j) = d(\widetilde{p_1}(i,j),{p_1}(i,j))  
\end{equation}

where $d(\widetilde{p_1}(i,j),{p_1}(i,j))$ is defined as the distance between the two probability distributions. Replacing $d(\cdot,\cdot)$ by KL-divergence, the objective function reduces to

\begin{equation}
  O_1 = -\sum_{(i,j)\in E}O_1(i,j)  
\end{equation}

\begin{equation}
  O_1 = -\sum_{(i,j)\in E}w_{ij}\log p_1(v_i,v_j)  
\end{equation}

\subsubsection{Network embedding using second order proximity:}

The second order proximity of two nodes $(v_i,v_j)$ measures the similarity of the neighbourhood network structures of nodes $(v_i,v_j)$. This measure is applicable for both directed and undirected graphs. Hence our objective, in this case, is to look at the vertex and the ``context'' of the vertex which can also be related to the distribution of neighbours of the given vertex. Hence for each edge $(v_i,v_j)$ the probability of ``context'' is defined by

\begin{equation}
  p_2(v_j|v_i) = \frac{exp(\vec{u'}^T_j \cdot \vec{u}_i)}{\sum_{k=1}^{|V|}exp(\vec{u'}^T_k \cdot \vec{u}_i)}  
\end{equation}

Where $| V |$ is the number of vertices. As mentioned before, the second order proximity assumes that vertices with similar distribution over the contexts as similar vertices. To maintain the second order proximity, the similarity distance between the contexts $p_2(\cdot | v_i)$ represented in the low dimensional vector space and the empirical distribution $\widetilde{p_2}(\cdot | j)$ must be optimized. Hence our objective function ($O_2$) in this case is

\begin{equation}
  O_2 = \sum_{v_i \in V}\lambda_id(\widetilde{p_2}(\cdot|v_i),p_2(\cdot|v_i))  
\end{equation}

where $d(\cdot , \cdot)$ is the distance between two probability distributions, here the variable $\lambda_i$ is used to consider the importance of the vertex $v_i$ during the process of optimization. As defined in the previous case the empirical  distribution is defined as 

\begin{equation}
  \widetilde{p_2}(i,j) = \frac{w_{ij}}{d_i} \ \ \ and \ \ \ d_i = \sum_{k \in N(i)}^{}w_{ik}  
\end{equation}

$w_{ij}$ is the weight of edge $(v_i,v_j)$ and $d_i$ is defined as the out-degree of vertex and N(i) is the set of neighbours of $v_i$. Considering $\lambda_i = d_i$ for the purpose of simplicity, and replacing $d(\cdot , \cdot)$ with KL-divergence

\begin{equation}
  O_2 = -\sum_{(i,j)\in E}w_{ij}\log p_2(v_j|v_i)  
\end{equation}

\subsection{Model Optimization}

The approach of negative sampling proposed by Mikolov et al. \cite{mikolov2013distributed} is used to optimize the objective function which helps us to represent every vertex of the network graph in the low dimensional vector space. Hence the objective function simplifies to:

\begin{equation}
  \log \sigma(\vec{u'}^T_j \cdot \vec{u}_i)) + \sum_{i=1}^K E_{v_n  P_n(v)}{[\log \sigma(\vec{u'}^T_j \cdot \vec{u}_i)]}  
\end{equation}

where $\sigma(x) = 1/(1+exp(-x))$ is the sigmoid function. We use the stochastic gradient descent algorithm \cite{recht2011hogwild} for optimizing the objective function and we update the model parameters on a batch of edges. Thus after completion of the training process, we get the embeddings corresponding to each vertex. The gradient with respect to an embedding $\vec{u_i}$ of vertex $v_i$ will be:

\begin{equation}
  \frac{\partial O_1}{\partial \vec{u_i}} = w_{ij}. \frac{\partial \log(p_1)(v_i,v_j)}{\partial \vec{u_i}}  
\end{equation}

\begin{equation}
    \frac{\partial O_2}{\partial \vec{u_i}} = w_{ij}. \frac{\partial \log(p_2)(v_j|v_i)}{\partial \vec{u_i}}
\end{equation}

We learn the node embeddings ($\vec{u_i}$) by optimizing the objective function in both cases and call the embeddings as first order embeddings and second order embeddings respectively.The model is trained using the Tensorflow (\cite{abadi2016tensorflow}) library on a cuda GPU. Model is trained using RMS Propagation gradient descent algorithm with learning rate as 0.025, and we used a batch size as $128$, the number of batches = $300000$ and 300-dimensional embeddings. The code is made available on Github\footnote{\url{https://bit.ly/2I5hYNd}}, 300-dimensional emoji embeddings learned using the emoji co-occurrence network can also be accessed at this link.

\section{Experiments}
\label{sec:Experiments}
\subsection{Sentiment Analysis}

In this section, we report our accuracies obtained for the sentiment analysis task on the gold-standard dataset developed by Novak et al. \cite{novak2015sentiment}. Our experiments have achieved accuracies which outperform the current state-of-the-art results for sentiment analysis on the gold-standard dataset. The gold-standard dataset\footnote{\url{https://bit.ly/2pLaKVZ}} consists of 64599 manually labelled tweets classified into positive, negative, neutral sentiment. The dataset is divided into training set that consists 51679 tweets, 9405 out of which contain emoji and testing set that consists of 12920 tweets, 2295 out of which contain emoji. In both training and testing sets, 29\% are labelled as positive, 25\% are labelled as negative, and 46\% are labelled as neural. We use the pre-trained FastText word embeddings\footnote{https://bit.ly/2FMTB4N} \cite{mikolov2018advances} to embed words into a low dimensional vector space. We calculate the bag of words vector for each tweet and then use this vector as a feature to train a support vector machine and a random forest model on the training set, and evaluate the accuracies obtained for classification task on whole testing dataset consisting of 12920 tweets. The accuracies obtained for classification task using the first order embeddings surpass the current state-of-the-art~\cite{wijeratne2017semantics} results.

\begin{table}[ht]
\begin{center}
\caption{Accuracy of Sentiment Analysis task}
\label{sec:sentimenttable}
\bigskip

\begin{tabular}{| >{\centering\arraybackslash}m{0.8in} | >{\centering\arraybackslash}m{0.8in} | >{\centering\arraybackslash}m{0.8in} |}
\hline
Word Embeddings & Classification accuracy using RF & Classification accuracy using SVM\\
\hline
State-of-the-art results & 60.7 & 63.6 \\
\hline
First Order Embedding & \textbf{62.1} & \textbf{65.2} \\
\hline
Second Order Embedding & 58.7 & 61.9 \\
\hline
\end{tabular}
\end{center}
\end{table}

\subsection{Emoji Similarity}

Emoji similarity\footnote{Our main objective is not to address the emoji similarity task. Our main objective is to demonstrate the usefulness of our emoji embeddings for sentiment analysis task.} is one of the important challenges which should be addressed for the development of emoji keyboards since the current emoji keyboard consists of 2666 emojis, and the complete list cannot be accommodated in a small screen. These emoji embeddings learned using the emoji co-occurrence network graph could be helpful to calculate the similarity between emojis using cosine distance as the similarity measure and group emojis which have high similarity values. This grouping of emojis can decrease the number of distinct emojis and helps us accommodate this grouped emojis on a small screen. In this section, we report the emoji similarity values found considering the first order embeddings and second order embeddings.

We consider the cosine distance to be the similarity measure between two embeddings. Let $\vec{a}$ and $\vec{b}$ be two vectors which represent embeddings of emojis $e_1$ and $e_2$ respectively, the similarity measure between these two emojis ($e_1$ and $e_2$) is calculated as

\begin{equation}
  similarity(e_1,e_2) = \frac{\vec{a} \cdot \vec{b}}{|a|\cdot |b|}  
\end{equation}

\hyperref[sec:firstorder]{Table 5} and \hyperref[sec:secondorder]{Table 6} reports the most similar emojis found considering the first order embeddings and second order embeddings respectively.The observed results are explained in \hyperref[sec:discussion]{Section 6}.

\begin{table}[ht]
\begin{center}
\caption{Emoji Similarity Measured using first order embeddings}

\bigskip
\label{sec:firstorder}
\begin{tabular}{| >{\centering\arraybackslash}m{0.8in} | >{\centering\arraybackslash}m{0.8in} | >{\centering\arraybackslash}m{0.8in} |}
\hline
Emoji Pair & Similarity Measure & Semantic Similarity\\
\hline
(\includegraphics[height=1em]{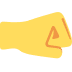},\includegraphics[height=1em]{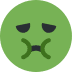}) & 0.921 & 0.442\\
(\includegraphics[height=1em]{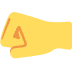},\includegraphics[height=1em]{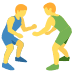}) & 0.916 & 0.598\\
(\includegraphics[height=1em]{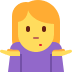},\includegraphics[height=1em]{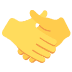}) & 0.911 & 0.623\\
(\includegraphics[height=1em]{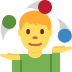},\includegraphics[height=1em]{Emojis/110.png}) & 0.909 & 0.546\\
(\includegraphics[height=1em]{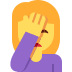},\includegraphics[height=1em]{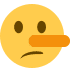}) & 0.856 & 0.723\\
(\includegraphics[height=1em]{Emojis/102.png},\includegraphics[height=1em]{Emojis/98.png}) & 0.889 & 0.702\\
(\includegraphics[height=1em]{Emojis/98.png},\includegraphics[height=1em]{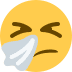}) & 0.881 & 0.737\\
\hline
\end{tabular}
\end{center}
\end{table}

\subsection{Analogical Reasoning}

The analogical reasoning task introduced by Mikolov et al. \cite{mikolov2013distributed}, defines the syntactic and semantic analogies. For example, consider the semantic analogy such as USA : Washington = India : ? where we fill the gap (represented by ``?'') by finding a word from the vocabulary whose embedding(represented by vec(x)) is closest to vec(Washington) - vec(USA) + vec(India). Here cosine distance is considered as the similarity measure between the two vectors.

\begin{table}[ht]
\begin{center}
\caption{Emoji Similarity Measured using second order embeddings}

\bigskip
\label{sec:secondorder}
\begin{tabular}{| >{\centering\arraybackslash}m{0.8in} | >{\centering\arraybackslash}m{0.8in} | >{\centering\arraybackslash}m{0.8in} |}
\hline
Emoji Pair & Similarity Measure & Semantic Similarity\\
\hline
(\includegraphics[height=1em]{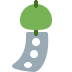},\includegraphics[height=1em]{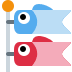}) & 0.646 & 0.662\\
(\includegraphics[height=1em]{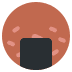},\includegraphics[height=1em]{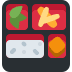}) & 0.606 & 0.598\\
(\includegraphics[height=1em]{Emojis/87.png},\includegraphics[height=1em]{Emojis/109.png}) & 0.596 & 0.623\\
(\includegraphics[height=1em]{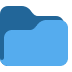},\includegraphics[height=1em]{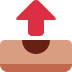}) & 0.556 & 0.622\\
(\includegraphics[height=1em]{Emojis/2290.png},\includegraphics[height=1em]{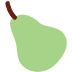}) & 0.546 & 0.916\\
(\includegraphics[height=1em]{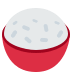},\includegraphics[height=1em]{Emojis/2266.png}) & 0.540 & 0.945\\
\hline
\end{tabular}
\end{center}
\end{table}

\begin{table}[ht]
\begin{center}
\caption{Spearman\textsc{\char13}s Rank Correlation Results}
\label{sec:spearman}
\bigskip

\begin{tabular}{| >{\centering\arraybackslash}m{1.6in} | >{\centering\arraybackslash}m{0.4in} |}
\hline
\textbf{Emoji Embeddings} & $\mathbf{\rho * 100}$\\
\hline
First Order Embeddings & \textbf{74}\\
\hline
Second Order Embedding & 66\\
\hline
\end{tabular}
\end{center}
\end{table}

\subsubsection{Emoji to emoji analogy}

We extrapolate the semantic analogy task introduced by Mikolov et al. \cite{mikolov2013distributed} in the context of emojis, by replacing words with emojis. Consider an emoji analogy, (\includegraphics[height=1em]{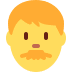} : \includegraphics[height=1em]{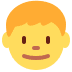}) = (\includegraphics[height=1em]{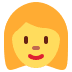} : ?), we fill the gap (represented by ``?'') by finding an emoji from the complete list of emojis whose embedding(represented by vec(x)) is closest to vec(\includegraphics[height=1em]{Emojis/1738.png}) - vec(\includegraphics[height=1em]{Emojis/1736.png}) + vec(\includegraphics[height=1em]{Emojis/1735.png}). \hyperref[sec:emoji2emoji]{Table 8} reports some of the interesting analogies found using first order and second order embeddings.

\begin{table}[ht]
\begin{center}
\caption{Emoji to Emoji Analogical Reasoning using Emoji Embeddings}

\bigskip
\label{sec:emoji2emoji}
\begin{tabular}{| >{\centering\arraybackslash}m{1in} | >{\centering\arraybackslash}m{1in} |}
\hline
First Emoji Pair & Second Emoji Pair \\
\hline
(\includegraphics[height=1em]{Emojis/1736.png} : \includegraphics[height=1em]{Emojis/1738.png}) & (\includegraphics[height=1em]{Emojis/1735.png} : \includegraphics[height=1em]{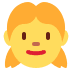}) \\
(\includegraphics[height=1em]{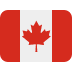} : \includegraphics[height=1em]{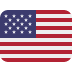}) & (\includegraphics[height=1em]{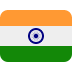} : \includegraphics[height=1em]{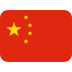}) \\
(\includegraphics[height=1em]{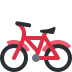} : \includegraphics[height=1em]{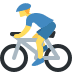}) & (\includegraphics[height=1em]{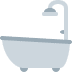} : \includegraphics[height=1em]{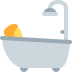}) \\
(\includegraphics[height=1em]{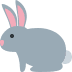} : \includegraphics[height=1em]{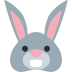}) & (\includegraphics[height=1em]{Emojis/1889.png} : \includegraphics[height=1em]{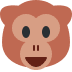}) \\
(\includegraphics[height=1em]{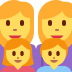} : \includegraphics[height=1em]{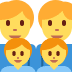}) & (\includegraphics[height=1em]{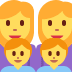} : \includegraphics[height=1em]{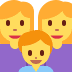}) \\

\hline
\end{tabular}
\end{center}
\end{table}

\section{Discussion}
\label{sec:discussion}

The high accuracy for classification task using the first order embedding model is due to the fact that all co-occurring emojis in a tweet possess the same sentiment feature, hence during classification these embeddings would increase the accuracy of the classification model. Consider the tweet, ``Who uses this emoji \includegraphics[height=1em]{Emojis/991.png}, I miss the one that had this  mouth \includegraphics[height=1em]{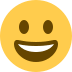} and these eyes \includegraphics[height=1em]{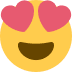} ! ... where did he go?! Why did he leave?!'' , in this tweet we observe the overall sentiment to be positive, and we also observe that all the emojis embedded in the tweet possess the same sentiment. Hence co-occurring emojis would be better attribute to learn emoji embeddings which can increase the accuracy of sentiment analysis and other related classification tasks.

We use the Spearman\textsc{\char13}s rank correlation coefficient to evaluate the emoji similarity ranks obtained using first order and second order embeddings learned using emoji co-occurrence network with the emoji similarity ranks of gold-standard dataset\footnote{\url{https://bit.ly/2GztSR2}}. \hyperref[sec:spearman]{Table 7} reports the Spearman\textsc{\char13}s correlation coefficient obtained by our emoji embeddings. According to the correlation coefficients the first emoji embeddings show a strong correlation ($ 0.6 < \rho < 0.79$).

The top 6 most similar emoji pairs observed considering the first order embeddings are reported in \hyperref[sec:firstorder]{Table 5}. As we see from \hyperref[sec:firstorder]{Table 5} the most similar emoji pair observed is (\includegraphics[height=1em]{Emojis/110.png},\includegraphics[height=1em]{Emojis/102.png}) with similarity measure of $0.921$. The usage of the first emoji \includegraphics[height=1em]{Emojis/110.png} would be in the context where the user wishes to express his dis-concern over certain issue through an act of hitting, the usage of the other emoji  \includegraphics[height=1em]{Emojis/102.png} would be in the context where the user wishes to express his dis-concern over certain issue through an expression of uneasiness. Hence the high similarity measure has sound even if we consider the context of use of the emojis. The results show that our embeddings give higher similarity measures than the semantic similarity\footnote{The semantic similarity is the similarity measure obtained using semantic embeddings developed by Wijeratne et al.} measure.

The top 6 most similar emoji pairs observed considering the second order embeddings are reported in \hyperref[sec:secondorder]{Table 6}. As we see from \hyperref[sec:secondorder]{Table 6} the most similar emoji pair observer is (\includegraphics[height=1em]{Emojis/2186.png},\includegraphics[height=1em]{Emojis/2187.png}) with similarity measure of $0.586$. The usage of the first emoji \includegraphics[height=1em]{Emojis/2186.png} would be in the context where the user wishes to generate a sound or ring a bell or in the context of celebration, the usage of the second emoji \includegraphics[height=1em]{Emojis/2187.png} would be in the context of celebration. EmojiNet lists ``celebration'' as a sense form for both the emojis, hence the observed similarity has sound even if we consider the context of use of this emojis.

\section{Future Work}

Usage of external knowledge has improved the accuracies of various natural language processing tasks and outperformed many state-of-the-art results. Jian et al. \cite{bian2014knowledge} have worked on leveraging external knowledge in learning word embeddings which gave better accuracies in word similarity and word analogy tasks. The first set of examples in EmoSim508\footnote{https://bit.ly/2GztSR2} dataset look more convincing than the results in \hyperref[sec:firstorder]{Table 5} and \hyperref[sec:secondorder]{Table 6}; the reason being semantic knowledge helps to us compare the similarity between different emojis efficiently. Using Jian et al.\textsc{\char13}s work as a reference, we could work on incorporating external knowledge from EmojiNet to our network embedding model which might further improve the accuracies of sentiment analysis and emoji similarity tasks.

\section*{Acknowledgement}
We are grateful to Sanjaya Wijeratne and Amit Sheth for thought-provoking discussions on the topic. We acknowledge support from the Indian Institute of Technology Kharagpur. Any opinions, findings, and conclusions/recommendations expressed in this material are those of the author(s) and do not necessarily reflect the views of Indian Institute of Technology Kharagpur.

\bibliographystyle{alpha} 
\newcommand{\etalchar}[1]{$^{#1}$}

\end{document}